\documentclass[namedreferences]{solarphysics}
\usepackage[optionalrh]{spr-sola-addons} 
\usepackage{graphicx}        
\usepackage{color}           
\usepackage{url}             

\newcommand{\etal}{{\it et al.}}

\newcommand{\dg}{$^\mathrm{o}$\,}

\begin{document}

\begin{article}
\begin{opening}

\title{Acoustic Mode Frequencies of the Sun during the Minimum Phase between Solar Cycles 23 and 24}
\author{S.~C. ~\surname{Tripathy}\sep
        K.~\surname{Jain}\sep
        F.~\surname{Hill}      
       }
\runningauthor{Tripathy \etal}
\runningtitle{Acoustic Mode Frequencies during Solar Minimum }

    \institute{$^{1}$ National Solar Observatory, Tucson, AZ 85719, USA\\
                     email: \url{stripathy@nso.edu} email: \url{kjain@nso.edu} 
 email: \url{fhill@nso.edu} \\
             }
\begin{abstract}
We investigate the spatial and temporal variations of the high-degree mode frequencies calculated over localized regions of the Sun during the extended minimum phase between solar cycles 23 and 24. The frequency shifts measured relative to the spatial average over the solar disk indicate that  the correlation between the frequency shift and magnetic field strength
during the low-activity phase is weak. The disk-averaged  frequency shifts computed relative to a minimal activity period also reveal a moderate correlation with different activity indices, with a maximum linear correlation of about 72\%. From the investigation of the frequency shifts at different latitudinal bands, we do not find a consensus period for the onset of  solar cycle 24. The frequency shifts corresponding to  most of the latitudes in the northern hemisphere and 30{\dg} south of the equator indicate the minimum epoch to be February 2008, which is earlier than inferred from solar activity indices. 
\end{abstract}
\keywords{Data analysis; Helioseismology; Solar cycle}
\end{opening}

\section{Introduction}
     \label{S-Introduction} 
The analysis of continuous helioseismic data over solar cycle 23 has produced many interesting results related to the cyclic variation of the oscillation frequencies and other mode parameters.  Most of the studies are confined to the low- and intermediate-degree modes (\opencite{jain11}; and references therein) which reinforces the idea that the mode frequencies change with solar activity level. However, the period of extended minimum between cycles 23 and 24 produced some surprises, most notably a disagreement between the low- and intermediate-degree modes regarding the onset of cycle 24. Based on  data from the { Global Oscillation at Low Frequency} (GOLF) instrument on-board the {\it Solar and Heliospheric Observatory} (SOHO), \inlinecite{salabert09} pointed out that cycle 24 started in late 2007. This was later substantiated by the analysis of {Birmingham Solar Oscillation Network} (BiSON) data \cite{bison}. But the analysis of intermediate-degree mode frequencies from {Global Oscillation Network Group} (GONG) and {Michelson Doppler Imager} (MDI) did not confirm the early onset \cite{jain10c}. Instead the occurrence of the minimum was shown to be around the middle of 2008 for MDI data and the end of 2008 for GONG data \cite{tripathy10}. The latter period approximately coincided with the minimum period observed in solar activity indices. A more detailed investigation of low- and intermediate-degree modes from GONG data, to eliminate the instrument dependence, demonstrated that the extended minimum period as seen in oscillation frequencies consist of two minima depending on the degree of the modes. The frequencies when analyzed as a function of the latitude also showed two minima at high latitude in contrast to a single minimum at mid--low latitudes \cite{jain11}. 

In the context of the extended minimum, we present here results obtained from the analysis of the high-degree mode frequencies  in the degree range of  $180 \le \ell \le 1000$ obtained using a technique of local helioseismology. With the use of a limited data set during the ascending phase of the solar cycle, \inlinecite{hindman} have explored the behavior of such high-degree modes and noted that the frequency variations are spatially as well as temporally associated with 
active regions. The analysis of high-degree global modes \cite{rhodes02}  showed that the frequency shifts of {\it p}-modes changed sign from positive to negative at frequencies near the acoustic cut-off and confirmed the earlier results   obtained from the analysis of intermediate-degree modes (\opencite{ronan94}; \opencite{jeff98}).  The higher frequencies of the high-degree modes were further found to be 
anti-correlated with solar activity (\opencite{rhodes03}; \opencite{rose03}). In the context  of local helioseismology, \inlinecite{howe08} studied frequencies of high-degree modes over localized regions of the Sun and found that the frequencies of modes beyond the acoustic cutoff have negative correlation with the local surface magnetic-field strength.
  The analysis of modes
$\ell \le 1000$ obtained from the global analysis during 1996--2008 \cite{rhodes11} not only confirmed the earlier results
but also indicated a further reversal around the frequency of approximately 6950 $\mu$Hz {\it i.e.} the 
frequencies were found to be positively correlated with the changes in activity indices. This investigation also pointed out that the frequencies where these reversals occur 
change with the mean level of activity. \inlinecite{christina11} has also investigated 
the properties of the globally determined intermediate and high-degree  modes ($20 \le \ell \le 900$) from MDI and reports evidence of a quadratic relation between the frequency shifts and 10.7 cm radio flux, $F_{10.7}$. 

Since the high-degree global oscillation modes cannot distinguish between the northern and southern hemispheres and are longitudinally averaged, we use the high-degree
modes obtained through the ring-diagram technique to examine their behavior at different latitudes during the extended minimum phase and to record the onset time of solar cycle 24.  We also explore the correlation between the frequency shifts and the strength of the magnetic activity during this phase. Preliminary results from a similar investigation but involving only a few Carrington rotations were reported by \citeauthor{sushant10c} (\citeyear{sushant10c},\citeyear{sushant11a}). The  correlation analysis is further extended to include other activity indices, {\it viz.} $F_{10.7}$ and the international sunspot number $R_{\rm I}$. 

----------

\begin{figure}    
   \centerline{\includegraphics[width=12cm,clip=]{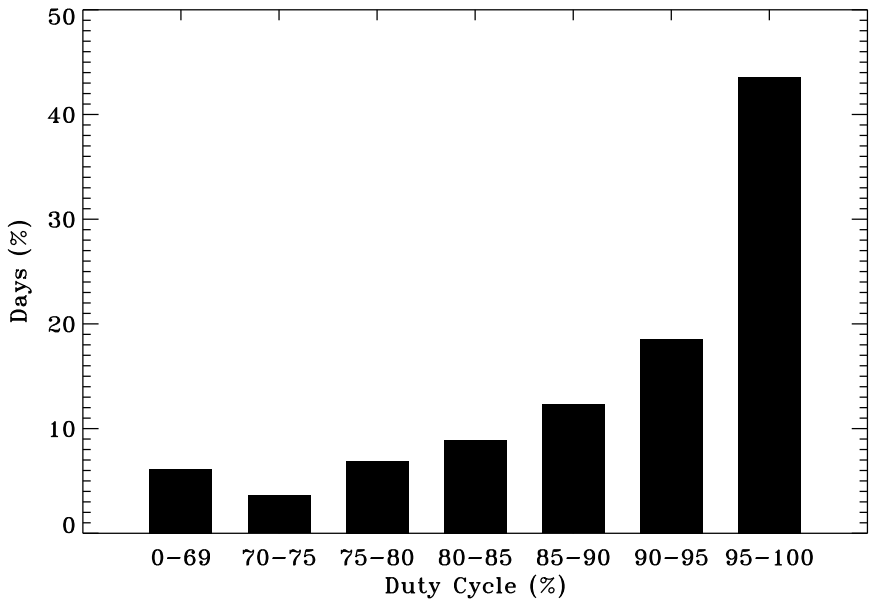}
              }
              \caption{Distribution of the duty cycle  of GONG ring-data between 26 April 2007 and  7 August 2010.} 
  \label{F-duty}
   \end{figure}
\section{Data and Analysis} 
We use the ring-diagram technique \cite{hill88} to calculate the high-degree mode frequencies in the range of 180 $\le \ell \le 1000$ and examine the spatial as well as temporal frequency shifts during the extended minimum phase covering Carrington rotations (CR) 2056  to 2099 starting with 26 April 2007 and ending with 7 August 2010. The mode frequencies corresponding to the 189 dense-pack overlapping tiles  are extracted from the GONG website {\url{http://gong.nso.edu}}.   
Each region covers an area of  16{\dg} $\times$ 16{\dg} in heliographic latitude and longitude and is tracked for a period of 1664 min (hereinafter one ring-day) using a model solar velocity \cite{snod84}. The centers of the regions are separated by 7.5{\dg} in latitude and longitude and extend to roughly 52.5{\dg} from the disk center, covering a substantial portion of the solar disk.  On the time domain, successive ring days are overlapped between 27 and 30 min. Each tracked region is then apodized with a circular function and converted to a three-dimensional power spectra by applying a FFT in both spatial and temporal directions \cite{corbard03}. The resulting power spectra are fitted with a Lorentzian profile model to obtain the mode parameters \cite{haber00}. The effective duty cycle of the data considered here ranges from 32\% to 100\% with a mean of about 90\%. Figure~\ref{F-duty} shows the distribution of duty cycle. For this analysis, we included only data for which the duty cycle exceeds  70\%, which yielded usable data for 991 out of 1056 ring-days and close to 180000 individual tiles. 

\begin{figure}    
   \centerline{\includegraphics[width=6.10cm,clip=]{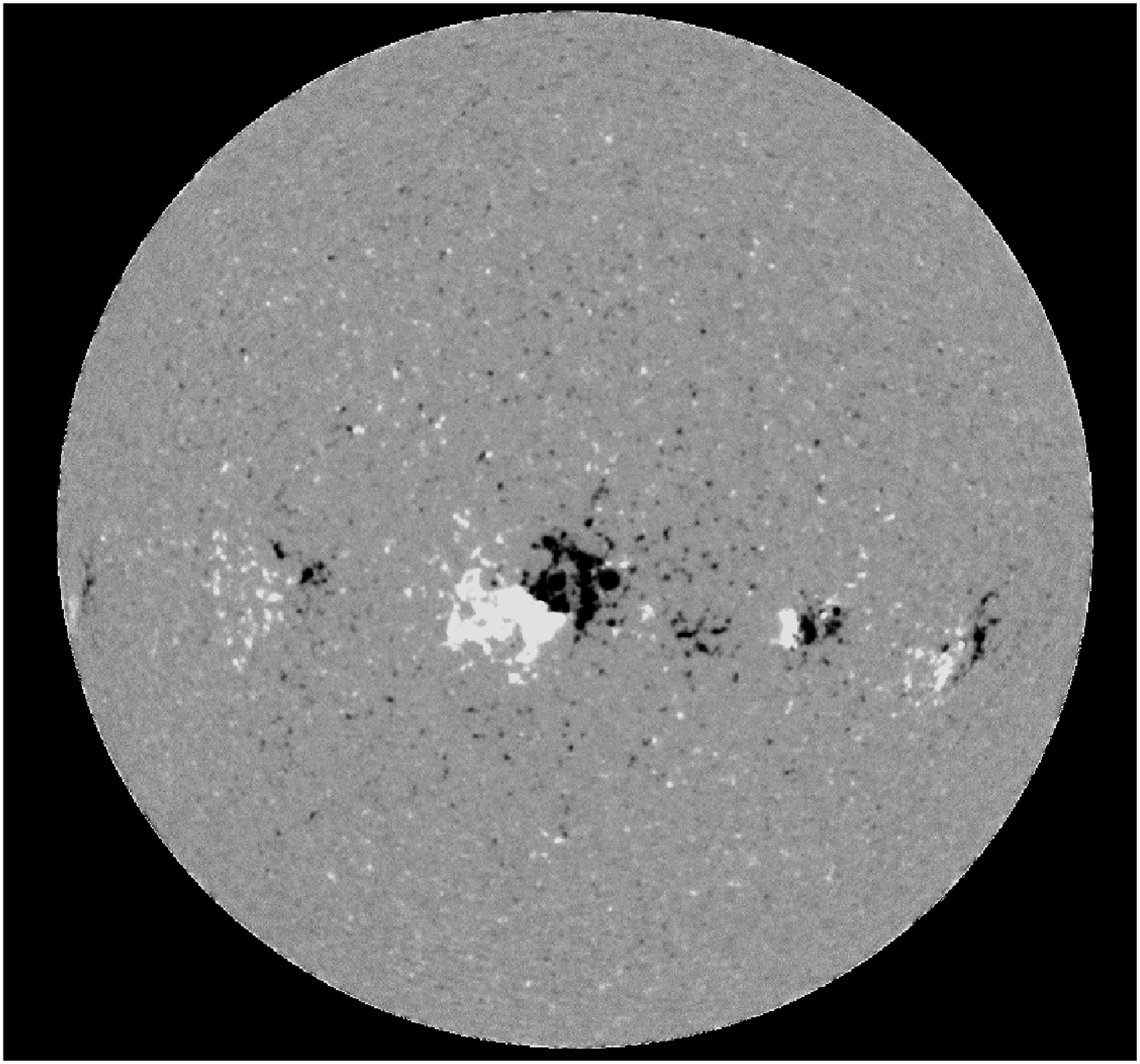}\\
              \includegraphics[width=5.7cm,clip=]{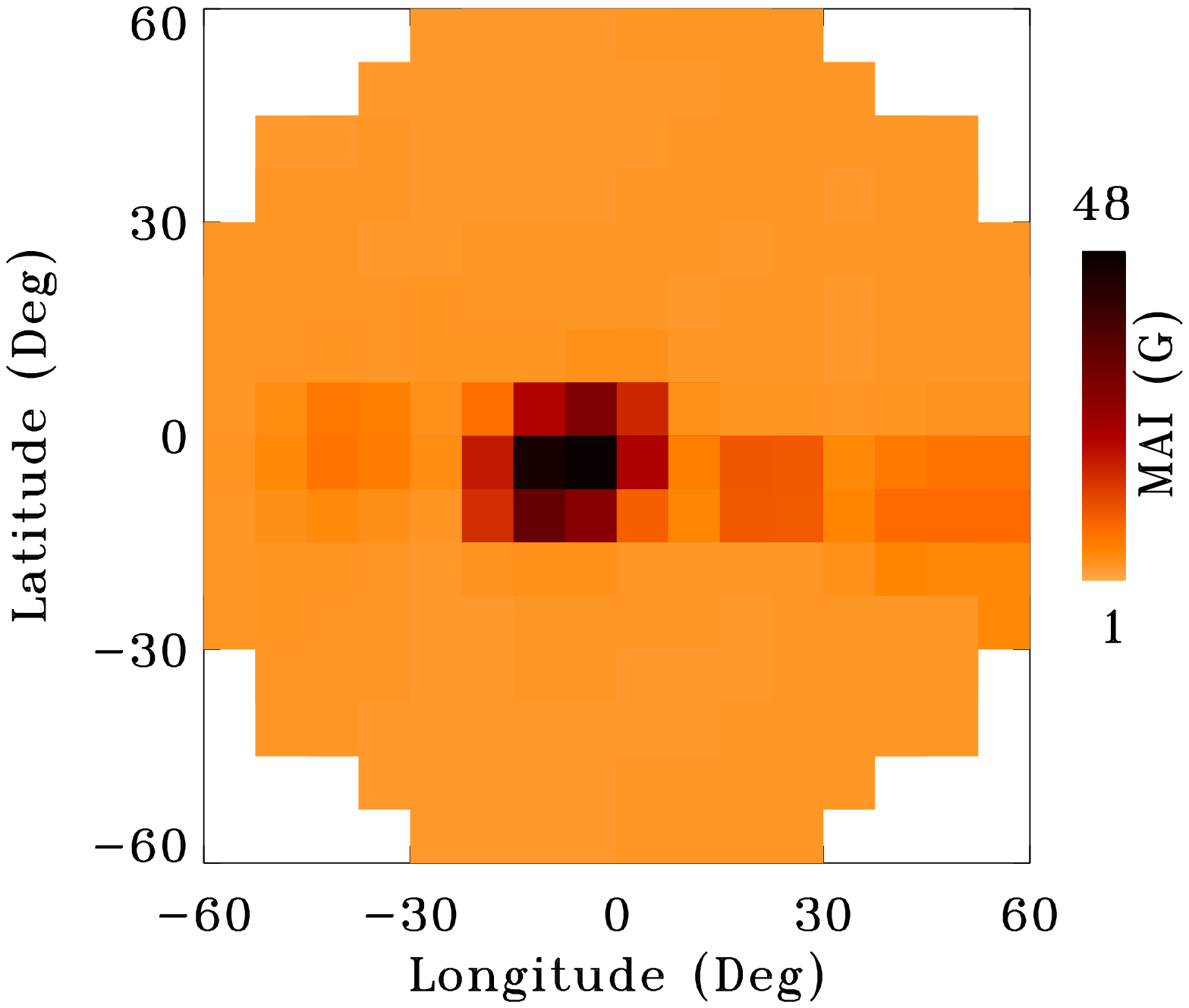}
              }
              \caption{GONG full-disk magnetogram taken at 13:57:16 UT, corresponding to the mid-time of the ring-day of 7 June 2007, is shown in the left panel. The MAI map of the same ring-day covering $\pm$ 60{\dg} in latitude and longitude is shown in the right panel.}
  \label{F-mag}
   \end{figure}
We also determine the level of local magnetic activity associated with each tile   by calculating a magnetic activity index (MAI) from the GONG magnetograms. In this study, we use magnetograms sampled every 32 min,   
mapped and tracked in the same way as the Dopplergrams. The remapped values are subjected to the same spatial apodization used in preparing the Doppler power spectra, so that both of them have the identical spatial extent. 
All of the pixels, with a field strength higher than  2.5 times the noise level of a given  pixel in GONG magnetograms, are then averaged to represent the MAI of the tile. Figure~\ref{F-mag} shows an example of the full-disk GONG magnetogram and the corresponding MAI map covering $\pm$ 60{\dg} in latitude and longitude for the ring-day 7 June 2007. The correspondence between the two is easily visible in the figure.

\begin{figure}    
   \centerline{\includegraphics[width=12cm,clip=]{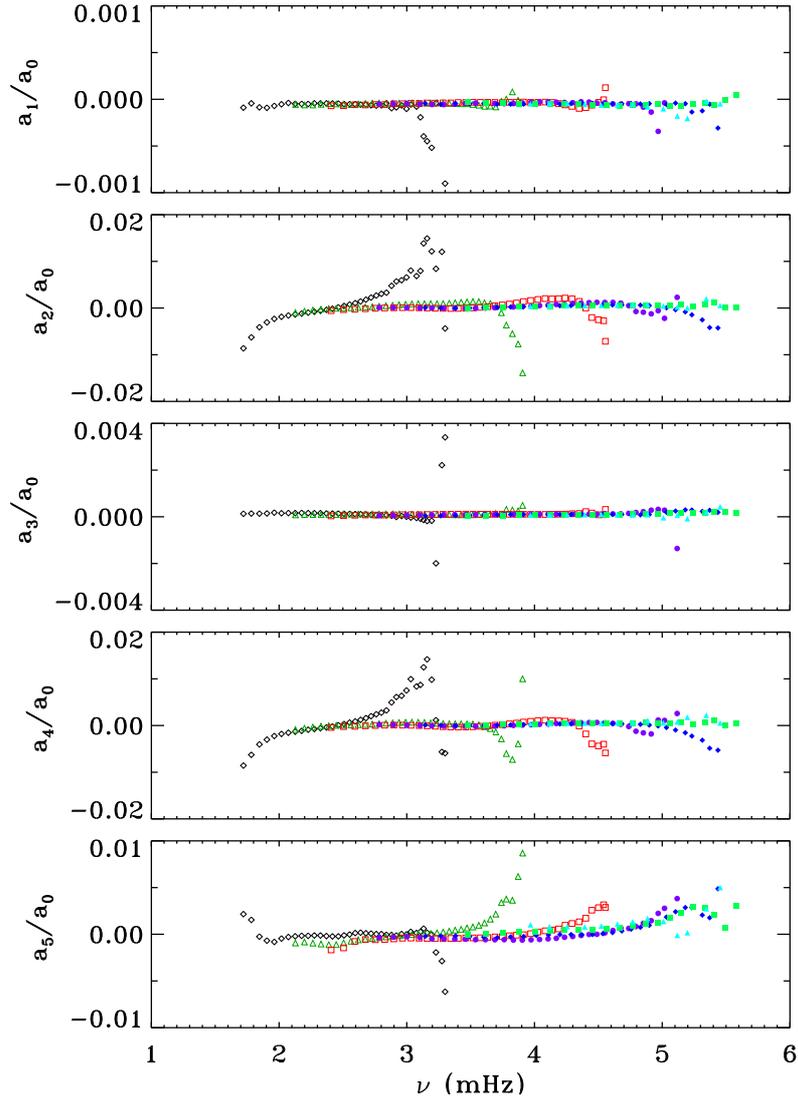}
              }
              \caption{Coefficients of the quadratic fit of frequencies to the  distance from the disk center and the duty cycle, divided by the constant term $a_0$. The symbols denote different $n$ values:  open diamond: $n$ = 0, open triangle: $n$ = 1, open square: $n$ = 2, filled circle: $n$ = 3, filled diamond: $n$ = 4, filled square: $n$ = 5, and  filled triangle: $n$ = 6.} 
  \label{F-coeff}
   \end{figure}
 
It has been shown earlier that the mode parameters  measured by the ring-diagram analysis are influenced by 
the foreshortening. There has also been evidence that the duty cycle of the observation introduces bias in the measured 
quantities. Therefore, it is necessary to correct the frequencies for the duty cycle and geometric effects.  Following  \inlinecite{howe04a}, we model the effects of position of the disk as a two-dimensional function of the distance from the disk center $\rho$ without the cross-terms, combined
with a linear dependence on the duty cycle $f(t)$ of the observations, 

\begin{figure}    
   \centerline{\includegraphics[width=12cm,clip=]{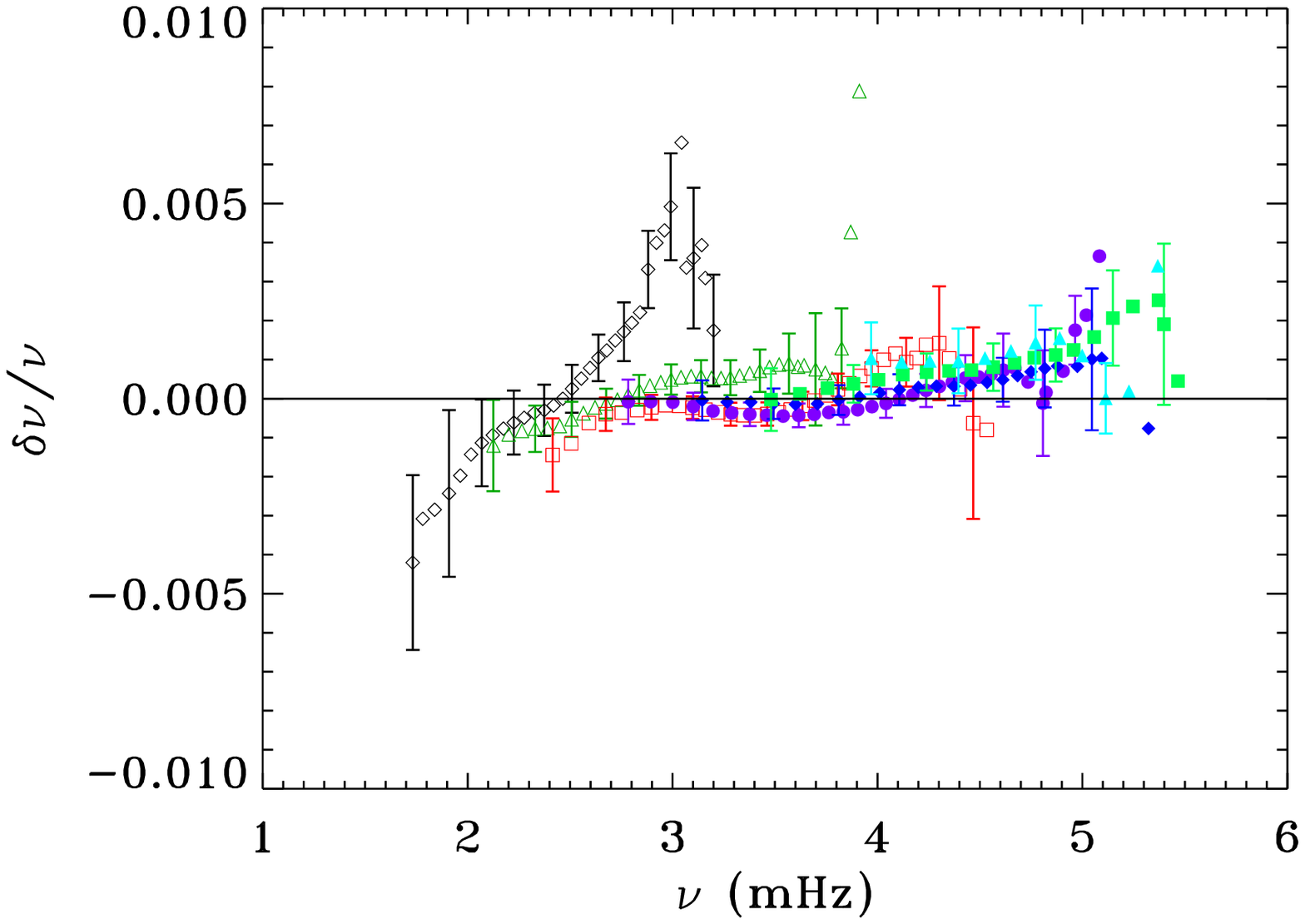}}
              \caption{Frequency differences between the corrected and fitted frequencies for a tile centered at a latitude of 52.5{\dg} north of the equator and  22.5{\dg} west of central meridian as a function of frequency. Errors for every third point are only shown. The symbols
denote different $n$ values open diamond: $n$ = 0, open triangle: $n$ = 1, open square: $n$ = 2, filled circle: $n$ = 3, filled diamond: $n$ = 4, filled square: $n$ = 5, and  filled triangle: $n$ = 6.} 
  \label{F-fdiff}
   \end{figure}

\begin{equation}
\nu(\rho_x,\rho_y,t)=a_0+a_1\rho_x+a_2\rho_x^2+a_3\rho_y+a_4\rho_y^2+a_5f(t)
\end{equation}

\noindent where   $\rho_x$ and $\rho_y$ are the longitudinal and latitudinal component of 
 $\rho$, and $a_i$ are coefficients determined by fitting.   In order to minimize 
the effect of the magnetic field on the fitting, we use the frequencies over a two year quiet period between 1 September  2007 and 31 August 2009 and evaluated the $a_i$s for each ($n$, $\ell$) mode. The corrected frequency of each mode
is then used for the analysis. Figure~\ref{F-coeff} illustrates the variations of the coefficients
as a function of frequency.   The linear terms are an order of magnitude smaller than the quadratic terms, but not insignificant.   Figure~\ref{F-fdiff} shows an example of the frequency differences at the extreme disk position 
within the 189 tiles located at a latitude of 52.5{\dg} north of the equator and 22.5{\dg} west of central meridian corresponding to 10 June 2008. 

\begin{figure}    
   \centerline{\includegraphics[width=12cm,clip=]{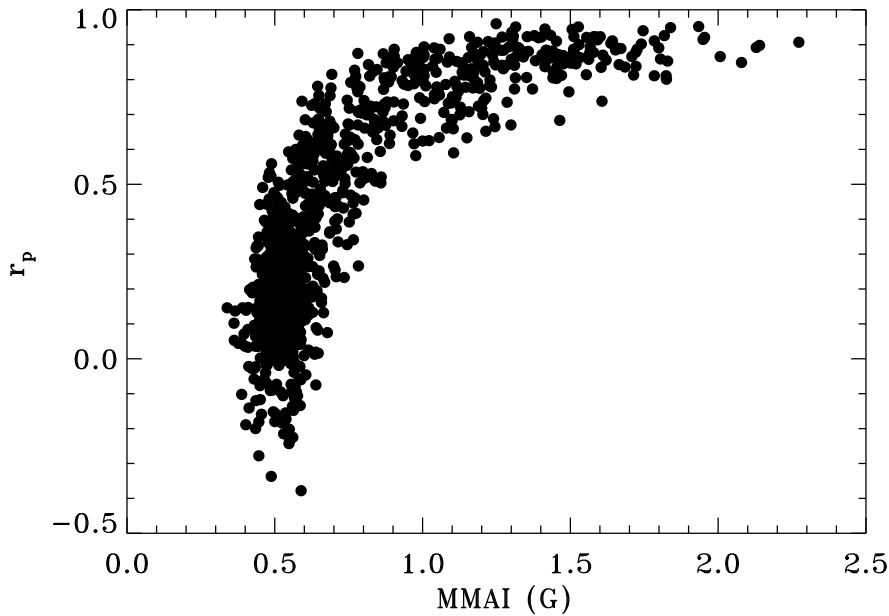}}
              \caption{Linear correlation coefficient $r_{\rm p}$ between the measurement of $\delta\nu_{\rm s}$ and MAI of 189 tiles 
for 991 ring-days.}
  \label{F-spatial}
   \end{figure}

\section{Results and Discussions}

\subsection{Spatial Variation}\label{S-spatial}
In order to explore the spatial variation of the frequencies and its correlation with the
 MAI of the corresponding tile, we calculate a mean spatial frequency shift for each ring-day. First, the frequency difference of each mode ($\delta\nu_{n,\ell}$)  
 is calculated with respect to a reference frequency which is the error-weighted average of that particular mode present 
in 189 dense-pack tiles. This difference is then averaged, weighted by the mean error, over all the modes present in the tile to estimate the spatial frequency shift ($\delta\nu_{\rm s}$) one value for each tile according to 
\begin{equation}
\delta\nu_{\rm s}\;=\,\sum_{n,\ell}\frac{\delta\nu_{n,\ell}}{\sigma_{n,\ell}^2}
/\sum_{n,\ell}\frac{1}{\sigma_{n,\ell}^2} . 
\end{equation}

As noted in \inlinecite{hindman},
the mean frequency shifts would have been zero but for the presence of activity on the solar disk.  Examples of the frequency shifts and coeval MAIs over 189 tiles  for two different phases of the solar cycle 23 are presented by  \inlinecite{sushant11a}. Here we concentrate on Pearson's linear correlation coefficient ($r_{\rm p}$) obtained between 189 $\delta\nu_{\rm s}$ and MAI for each ring-day.
The resultant correlation coefficients  as a function of the  mean magnetic activity index (MMAI) for 991 ring-days 
are shown in Figure~\ref{F-spatial}, where MMAI is the mean MAI over the 189 dense-pack tiles.         
  As shown in earlier studies, mostly confined to active regions \cite{rajaguru01} or high-activity periods \cite{howe04b}, the shifts, in general, have a higher correlation at higher MAIs.  However, Figure~\ref{F-spatial} indicates that tiles with small MAIs do not correlate well with the shifts,  indicating that the frequency shifts act as a tracer of magnetic field only if the MMAI is above a threshold value.  We also notice that during the lowest activity period, $\delta\nu$ is anti-correlated with MMAI. Although the values of the coefficients are not significant, it is worth mentioning that a similar result for the low- \cite{salabert09} and intermediate-degree  \cite{tripathy10} global-mode frequencies have been reported earlier.   
\begin{figure}    
   \centerline{\includegraphics[width=12cm,clip=]{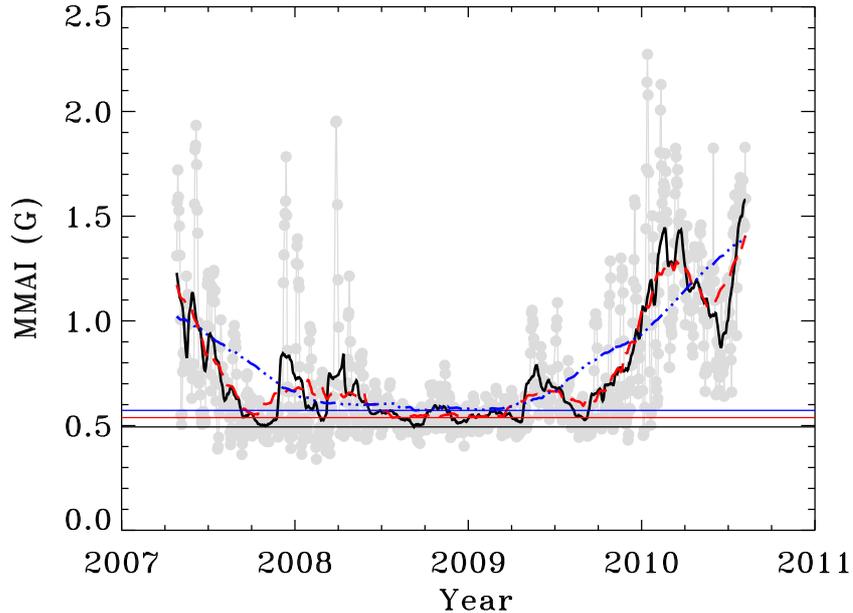}}
              \caption{Temporal evolution of mean magnetic activity index (MMAI). The black (solid), red (dashed) and blue (dashed--dot--dot--dot) lines represent the running mean over 23, 91, and 365 points. The horizontal lines indicate the position of the minimum epochs.}
   \label{F-diskavg}
   \end{figure}
\subsection{Temporal Variation}

Figure~\ref{F-diskavg} displays the temporal evolution of MMAI and  indicates 
that the extended minimum phase in solar activity occurred between 2007 and 2010 with the deepest phase 
lasting for  one year between mid-2008 and mid-2009. In order to eliminate the random fluctuations
produced by daily variations, we  calculate a 23-point running mean that approximately corresponds to an average over a Carrington rotation period. This is shown by the solid line in the figure. Although, most of the random fluctuations are reduced, we still notice rises and dips throughout the minimum phase. These fluctuations have been 
interpreted as a quasi-biennial signal  in global low-degree modes \cite{bison} and its origin is suspected to be near the shear layer just beneath the surface.   

From the smoothed data, it is easy to estimate the epoch of the solar minimum which is defined as the lowest value in the data set and found to be around September 2008 in agreement with the epoch of the minimum associated with other activity indices. However,  a value of similar magnitude as the minimum seen during September 2008 is also noticed during the third quarter of 2007, hinting at the possibility of existence of a double minimum. To test if the position of the minimum changes with the number of points used for smoothing, we additionally smooth the data with 91 (dashed line) and 365 (dash--dot--dot--dot line) points and find that the position of the minimum does not change significantly. Except for the data smoothed with 365 points, which probably represents  over-smoothing, we find the signature of  double minima in MMAI, the 2007 period coinciding with the minimum epoch inferred from the analysis of the low-degree modes \cite{salabert09} and the 2008 period with  intermediate-degree modes \cite{tripathy10}.

In order to characterize the temporal evolution of the mode frequencies,  we calculate the mean frequency shift ($\delta\nu$) 
using a formula which is analogous to the frequency shifts calculated for the global modes, 
\begin{equation}
\delta\nu\;=\,\sum_{n,\ell}\frac{\sum_{i=1}^{189}\,\delta\nu_{n,\ell}(t)}{\sigma_{n,\ell}^2}
/\sum_{n,\ell}\frac{1}{\sigma_{n,\ell}^2}, 
\end{equation}
where $\delta\nu_{n,\ell}$ is  the frequency difference with respect to the frequencies corresponding 
to the ring-day of 11 May 2008. Since  $\delta\nu$ represents the frequency shift over a large area of the solar disk
and covers the activity belt, we consider these shifts to represent global shifts over each ring-day. 

\begin{figure}    
   \centerline{\includegraphics[width=12cm,clip=]{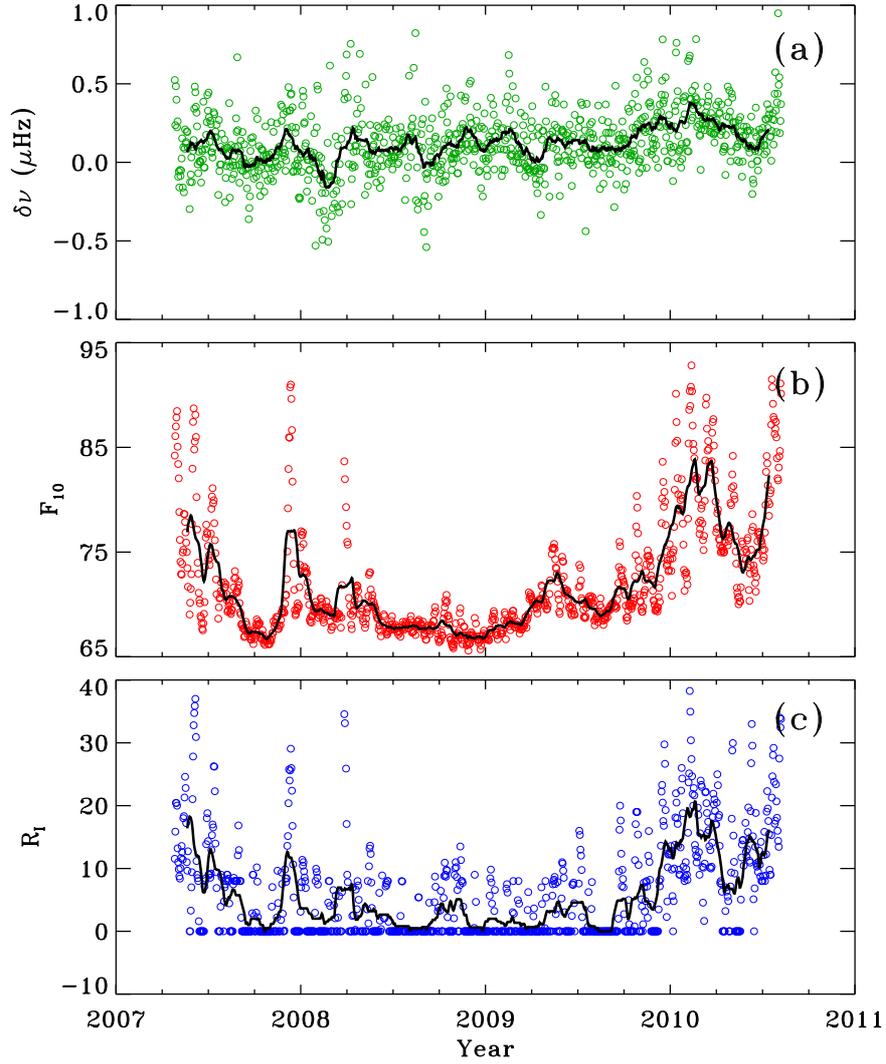}}
              \caption{Temporal evolution of (a) $\delta\nu$, (b)$F_{10.7}$, and (c) $R_{\rm I}$. The solid black line in each panel represents the running mean over 23 points.}
  \label{F-globalact}
   \end{figure}

The temporal variation 
of the calculated $\delta\nu$ and the associated running mean over 23 points is shown in Figure~\ref{F-globalact}. As before, 
we use the lowest value in smoothed frequency shifts to calculate the epoch of the solar minimum. It is evident that the minimum occurred in February 2008, roughly in agreement with the emergence of a sunspot group with the polarities of the new cycle during January 2008. However the appearance of several sunspots at low latitudes with cycle 23 polarities a few months later indicated otherwise \cite{philips}. Furthermore, the minimum seen in frequency shifts does not agree with the minimum epoch observed in MMAI hinting at a weak correlation between them.

Since  $\delta\nu$ represents disk-averaged quantity, we can further compute their relation with other global activity indices. In Figure~\ref{F-globalact}, we  show the temporal variation of  the international sunspot number ($R_{\rm I}$) and 10.7 cm radio flux ($F_{10.7}$) interpolated to the same temporal grid of frequency shifts.  The linear correlation coefficient between  $\delta\nu$ and different activity indices is presented in Table~\ref{T-global}. It is evident that the smoothed correlation is significantly higher than the raw correlation but in no case does the value exceed 72\% implying at best a  moderate correlation. As expected, the correlation is  higher between $\delta\nu$ and MMAI since these are calculated locally over the same tile while the other two indices reflect global measurements over the entire solar disk. 

\begin{table}
\caption{ Pearson linear ($r_{\rm p}$) and Spearman rank ($r_{\rm s}$) correlation statistics  between $\delta\nu$ and activity indices. The calculated probabilities of having null correlations in each case are smaller than $10^{-10}$.}
\label{T-global}
\begin{tabular}{lcccc}     
  \hline                   
Activity &\multicolumn{2}{c}{$r_{\rm p}$}&\multicolumn{2}{c}{$r_{\rm s}$}  \\     
        &Raw &Smooth &Raw &Smooth \\
  \hline
MMAI & 0.52&0.72 & 0.50&0.66 \\
$F_{10.7}$&0.41&0.68&0.36&0.58\\
$R_{\rm I}$& 0.39&0.65&0.38&0.57\\
  \hline
\end{tabular}
\end{table}
\begin{figure}[t]    
   \centerline{\includegraphics[width=12cm,clip=]{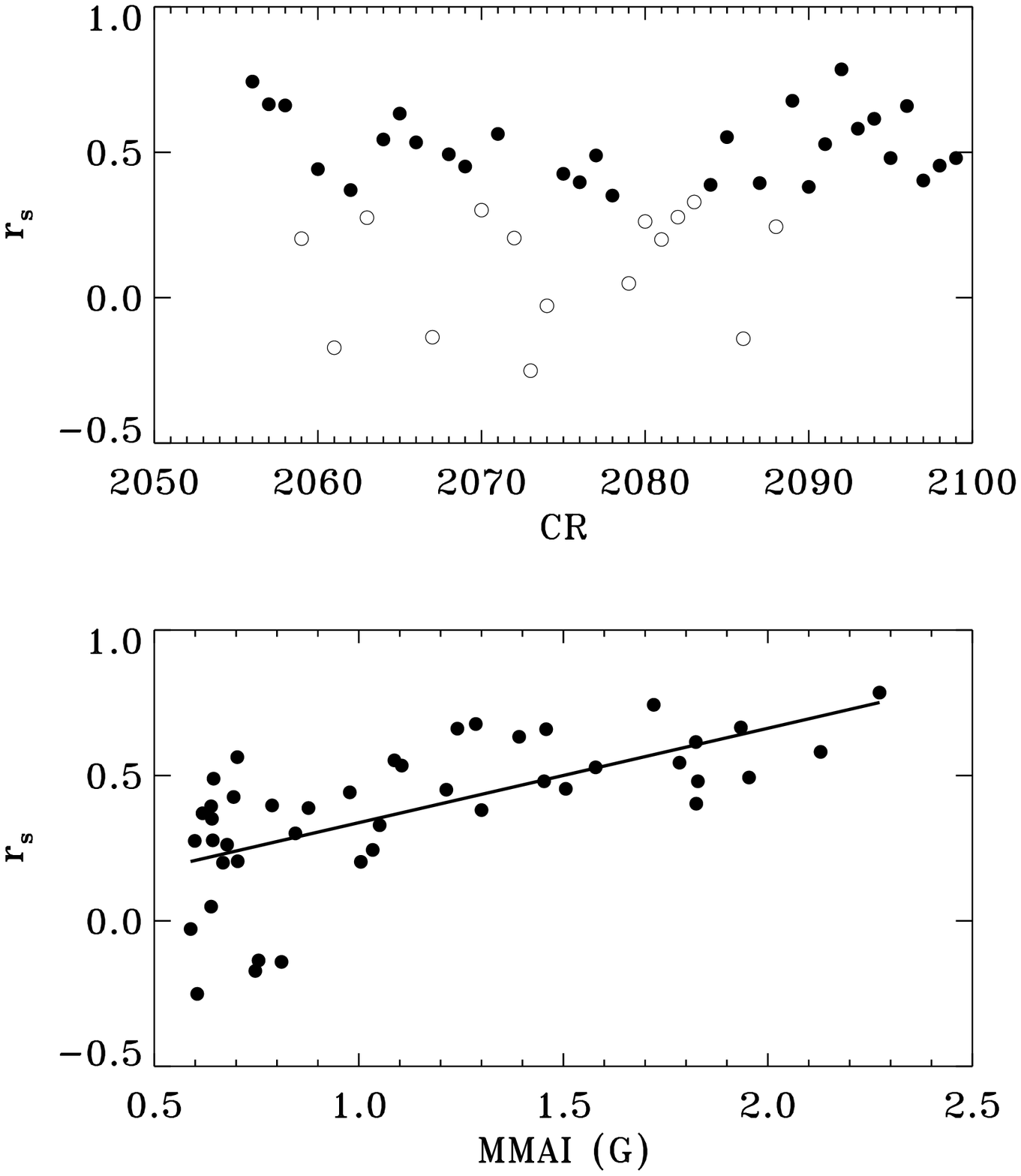}}
              \caption{Rank correlation coefficient ($r_{\rm s}$) between the  frequency shifts and MMAI calculated for each ring-day and sampled over each CR  (top panel) for different CRs and (b) as a function of MMAI.  The filled  (open) circles in the top panel represent the values where the two-sided significance is smaller (larger) than 0.1, respectively,  whereas in the bottom panel all the coefficients are shown by filled circles. The solid line in the bottom panel represents the linear fit. }
  \label{F-CRcor}
   \end{figure}
\subsection{Temporal Variation over CR} 
We further investigate the temporal variation of the frequency shifts for each CR separately, where the frequency difference 
of each multiplet ($\delta\nu_{n,\ell}(t)$) is calculated by subtracting an average frequency over the entire CR for each disk position. This difference is used in Equation~(3) to calculate the shift for each ring-day. It may be noted that not all CRs  include a uniform set of 24 ring-days since the data have been weeded for low duty-cycle values.   Figure~\ref{F-CRcor} shows the  rank correlation coefficients between the frequency shifts and MMAI as a function of time (top panel) and MMAI (bottom panel).  In the top panel, the filled (open) symbols refer to those values where the two-sided significance is smaller (larger) than 0.1 indicating significant (weak) correlation, respectively. Thus, the negative and few positive correlation coefficients with values less than about 0.5, shown by open symbols, are not significant. From the bottom panel, where all the coefficients are shown by filled symbols, it is evident that with the increase of MAI, the magnitude of the correlation coefficient slowly increases.  
However, the rank correlation between MMAI and $r_{\rm s}$ is about 65\% again implying a weak correlation.  The  variations in the rank correlation coefficients shown here are similar  to  Pearson's linear corrrelation coefficients shown in Figure~4 of \inlinecite{sushant11a} except for CR 2081. However, the values of the coefficients for the overlapping CRs 
between the two analyses cannot be compared since they represent coefficients calculated from two different correlation methods. 

\begin{figure}    
   \centerline{\includegraphics[width=120mm,clip=]{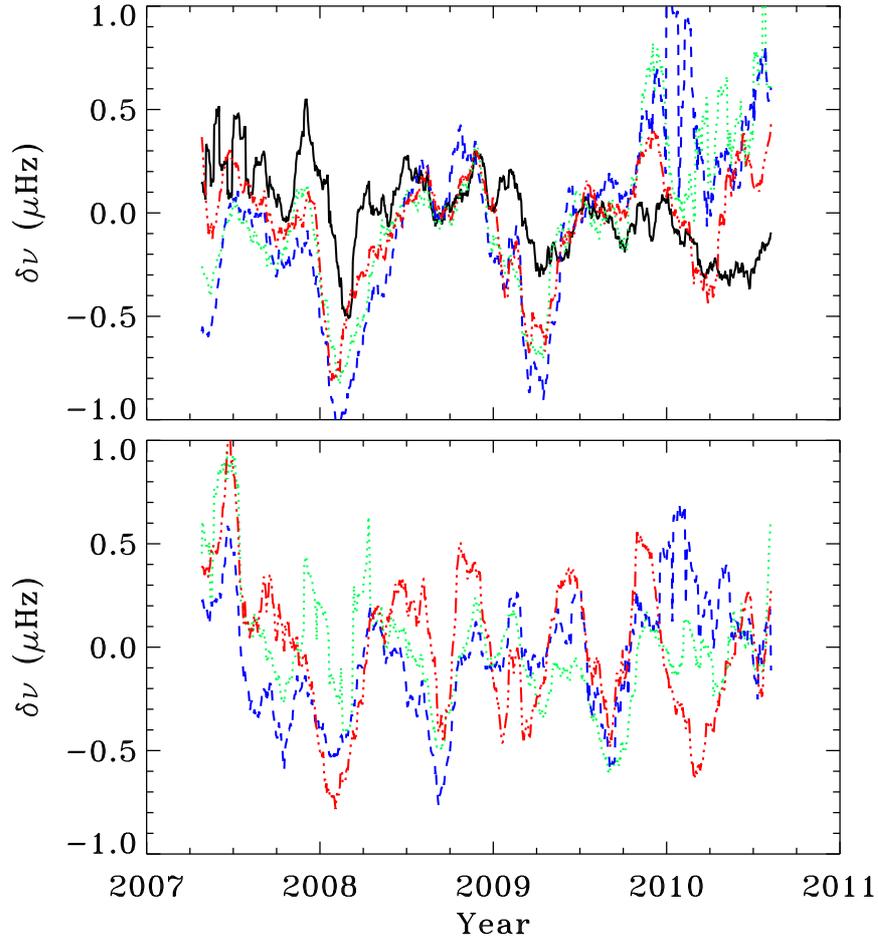}}
    \caption{Temporal variation of frequency shifts as a running mean over 23 points  at  selected latitudes on the central meridian; latitudes north of the equator including the equator (top panel) and latitudes south of the equator (bottom panel). Line styles have the following meaning: black (solid): equator, green (dotted): 15{\dg}, blue (dashed): 30{\dg}, and red (dash--three--dots): 45{\dg}.}
  \label{F-lat}
   \end{figure}
\subsection{Latitudinal Variation}
As mentioned earlier, there is no consensus regarding the epoch of the solar 
minimum based on the helioseismic data. Recent analysis of the global frequency shifts 
also indicates the possible existence of one or two minima at different latitudes \cite{jain11}. In order to find 
out if the mode frequencies calculated locally follow latitudinal patterns similar to global modes, we use Equation~(3) to calculate frequency shifts ($\delta\nu_{\rm lat}$) for different latitudes located on the central meridian. Here $\delta\nu_{n,\ell}$ is calculated  with respect to an average frequency at the given latitude on the central meridian over all the ring-days.  The running mean over 23 points of the resultant frequency shifts  for selected latitudes is displayed in Figure~\ref{F-lat}.  
 In the northern hemisphere (Figure~\ref{F-lat} (top)), the signature of the minimum can be clearly seen at two different epochs, the first minimum around February 2008 and the second around the first quarter of 2009. In case of the equator, we also notice a third dip around the middle of  2010 where the shifts are marginally smaller compared to the dip in 2009.  In the southern hemisphere, the picture is less clear as we notice  many dips associated with each of the curve. This may be a consequence of the fact that  during cycle 23, the southern hemisphere was more active compared to the northern hemisphere. As a result even a small amount of activity in the southern hemisphere would influence the mode frequencies and as a consequence our calculation of minimum based 
on the lowest value in the smoothed frequencies may not be the appropriate method. However, if we stick to this definition of the lowest value as the representation of the solar minimum, we find that the minimum in frequency shifts  at  15\dg,  30\dg, 
and  45\dg south of the equator   is around September 2009, September 2008, and February 2008, respectively.  In order to understand 
how closely $\delta\nu_{\rm lat}$ follows the magnetic activity at different latitudinal bands, we compute Pearson's linear and Spearman's rank  correlation coefficients between $\delta\nu_{\rm lat}$ and MAI of the tiles of the same latitude used for the calculation of frequency shifts. These coefficients along with the position of the minimum epoch for all latitude bands is given in Table~\ref{T-lat}. It is clear that the frequency shifts corresponding to different latitudes in the northern and southern hemisphere point to different epochs of the solar minimum with varying degree of correlation. In the equator and northern hemisphere, $\delta\nu_{\rm lat}$ of all the latitude bands except for the one located at 37.5\dg  indicate a minimum around February 2008 while in the southern hemisphere we do not have a consistent epoch of the minimum period, the significant deviation is seen 
in the tiles located at latitudes of 7.5\dg and 15\dg south of the equator. If we consider only those latitudes for which the 
correlation between  $\delta\nu_{\rm lat}$ and MAI is approximately 80\%, then we find that 50\% of the latitudes indicate the epoch of the minimum to be around February 2008 which is  consistent with the frequency shifts measured over the entire disk but earlier than inferred from MMAI and other solar activity indicators.    We also notice a small negative coefficient for the tiles located at 52.5\dg south of the equator  but the anti-correlation is not significant due to the low value of the correlation coefficient. Finally we note that the signature of the onset of solar cycle 24 as early as the third quarter of 2007 is not found in this analysis. 

\begin{table}
\caption{Pearson linear ($r_{\rm p}$) and Spearman rank ($r_{\rm s}$) correlation statistics at different latitudes between $\delta\nu_{\rm lat}$ and 
corresponding MAI and the time of occurrence of the minimum frequency shift.}
\label{T-lat}
\begin{tabular}{rccl}     
  \hline                   
Latitude &~~$r_{\rm p}$&~~$r_{\rm s}$& Minimum epoch  
      \\
  \hline
52.5{$^\mathrm{o}$} & ~~0.08 & ~~0.09 &February 2008\\
45.0{$^\mathrm{o}$} & ~~0.37 & ~~0.29 & February 2008\\
37.5{$^\mathrm{o}$} & ~~0.63 & ~~0.40 &March 2009\\
30.0{$^\mathrm{o}$} & ~~0.88 & ~~0.51&February 2008\\
22.5{$^\mathrm{o}$} & ~~0.85 & ~~0.57 &February 2008\\
15.0{$^\mathrm{o}$} & ~~0.83 & ~~0.43 &February 2008\\
7.5{$^\mathrm{o}$} & ~~0.57 & ~~0.32 &February 2008\\
0.0{$^\mathrm{o}$} & ~~0.78 & ~~0.45 &February 2008\\
$-$7.5{$^\mathrm{o}$} & ~~0.94 & ~~0.52 &April 2010\\
$-$15.0{$^\mathrm{o}$} & ~~0.85 & ~~0.49 &August 2009\\
$-$22.5{$^\mathrm{o}$}& ~~0.80 & ~~0.43  &September 2008\\
$-$30.0{$^\mathrm{o}$}& ~~0.79 & ~~0.38  &September 2008\\
$-$37.5{$^\mathrm{o}$} & ~~0.49 & ~~0.20 &February 2008\\
$-$45.0{$^\mathrm{o}$} & ~~0.04 & ~~0.02 &February 2008\\
$-$52.5{$^\mathrm{o}$} & $-$ 0.16~ &$-$ 0.17~ &February 2008 \\
  \hline
\end{tabular}
\end{table}

\section{Summary}

Using the ring-diagram technique applied to GONG data, we have examined the behavior of the high-degree modes during the extended minimum phase between cycles 23 and 24. Since the mode parameters measured by the ring-diagram technique are subject to foreshortening  and duty cycle effects, we  modeled the effect of the position of the disk as a two-dimensional function of the distance from the disk center combined with a linear dependence of the duty cycle and find that the correction factor is not significant in the 3~mHz band. 

Since it has been argued by \inlinecite{hindman} that active regions are responsible for the global-frequency shifts, we estimate the equivalent of a global frequency shift 
by computing an average over the local frequency shifts over the regions covered in dense-pack analysis. These shifts appear to be of the order of 1~$\mu$Hz during the minimum activity phase as opposed to 3~$\mu$Hz in the 3-mHz band in the 1998 MDI data  
deduced by \inlinecite{hindman}. This discrepancy may be attributed to the level of magnetic activity during the two periods. 

We have also investigated the epoch of the minimum phase as seen in the oscillation frequencies and find that the disk-averaged frequency shift indicates a minimum 
around February 2008 roughly in agreement with the emergence of a sunspot group with the polarities of the new cycle.  However, the times of minimum inferred from different latitudes do not, in general, agree with each other.  While the minimum at different latitudes in the northern hemisphere mostly agree the minimum to be around February 2008 with a lead time of about six months to the activity indicators, the minimum in the southern hemisphere shows a consistent time of February 2008 only for latitudes higher than 30\dg south of the equator where there is little or no activity. If we consider only those latitudes for which the correlation between  $\delta\nu_{\rm lat}$ and MAI is approximately 80\%, then we find that 50\% of the latitudes indicate the epoch of the minimum to be around February 2008 consistent with  the frequency shifts measured over the entire disk.

Since the magnetic-field strength of the Sun during the period of extended minimum phase is dominated by the quiet-sun magnetic field, we find that  both the spatial and temporal  shifts are weakly correlated with the surface magnetic activity. On this basis, we argue that  the shifts can-not be accounted for by the regions of observed component of the magnetic field alone but may be explained as a combined effect of the magnetic field and temperature changes \cite{kuhn01} or as an effect of a change in the acoustic cavity size \cite{dziem05}. The exact mechanism for these changes still remains an open question. 

\begin{acks}
We thank the reviewer for useful comments.  
This work utilizes data obtained by the Global Oscillation Network
Group (GONG) program, managed by the National Solar Observatory, which
is operated by AURA, Inc. under a cooperative agreement with the
National Science Foundation. The data were acquired by instruments
operated by the Big Bear Solar Observatory, High Altitude Observatory,
Learmonth Solar Observatory, Udaipur Solar Observatory, Instituto de
Astrof\'{\i}sico de Canarias, and Cerro Tololo Interamerican
Observatory. This work is partially supported by NASA grants 
NNG05HL41I and NNG08EI54I to the National Solar Observatory.
\end{acks}

\bibliographystyle{spr-mp-sola}
\bibliography{paper}

\end{article}
\end{document}